\documentclass[%
 reprint,
 superscriptaddress,
 amsmath,amssymb,
 aps,
 prl,
]{revtex4-1}

\usepackage[usenames,dvipsnames]{xcolor}
\usepackage{graphicx}
\usepackage{dcolumn}
\usepackage{bm}
\definecolor{myblue}{rgb}{0,0,1}
\usepackage[colorlinks=true,linkcolor=myblue,urlcolor=myblue,citecolor=myblue]{hyperref}

\usepackage{txfonts}

\let\vr\undefined

\newcommand{\vk}{{\bm{k}}}
\newcommand{\vr}{{\bm{r}}}

\begin{document}

\title{Spectral Functions of the Uniform Electron Gas via Coupled-Cluster Theory and
Comparison to the $GW$ and Related Approximations}

\author{James McClain}
\affiliation{Department of Chemistry, Princeton University, Princeton, NJ, USA}

\author{Johannes Lischner}
\affiliation{Department of Physics and Department of Materials, and the Thomas Young
Centre for Theory and Simulation of Materials, Imperial College, London, UK}
\affiliation{Department of Physics, University of California, Berkeley, CA, USA}
\affiliation{Materials Sciences Division, Lawrence Berkeley National Laboratory, Berkeley,
CA, USA}

\author{Thomas Watson}
\affiliation{Department of Chemistry, Princeton University, Princeton, NJ, USA}

\author{Devin A. Matthews}
\affiliation{Institute for Computational Engineering and Sciences, University of Texas at
Austin, Austin, TX, USA}

\author{Enrico~Ronca}
\affiliation{Department of Chemistry, Princeton University, Princeton, NJ, USA}

\author{Steven G. Louie}
\affiliation{Department of Physics, University of California, Berkeley, CA, USA}
\affiliation{Materials Sciences Division, Lawrence Berkeley National Laboratory, Berkeley,
CA, USA}

\author{Timothy C. Berkelbach}
\affiliation{Princeton Center for Theoretical Science, Princeton University, Princeton,
NJ, USA}

\author{Garnet Kin-Lic Chan}
\affiliation{Department of Chemistry, Princeton University, Princeton, NJ, USA}
\affiliation{Princeton Center for Theoretical Science, Princeton University, Princeton,
NJ, USA}

\date{\today}

\begin{abstract}
We use, for the first time, \textit{ab initio} coupled-cluster theory to compute the
spectral function of the uniform electron gas at a Wigner-Seitz radius of
$r_\mathrm{s}=4$. 
The coupled-cluster approximations we employ go significantly beyond the diagrammatic
content of state-of-the-art $GW$ theory. 
We compare our calculations extensively to $GW$ and $GW$-plus-cumulant theory,
illustrating the strengths and weaknesses of these methods in capturing the
quasiparticle and satellite features of the electron gas. 
Our accurate calculations further allow us to address the long-standing debate over the
occupied bandwidth of metallic sodium. 
Our findings indicate that the future application of coupled-cluster theory to condensed
phase material spectra is highly promising.
\end{abstract}

\maketitle

\textit{Introduction.}
Computing the electronic excitations and spectra of condensed phase systems with 
significant correlations from first-principles
continues to be a premier challenge in computational materials science.
Currently, a widely used approach is time-dependent many-body perturbation theory (MBPT).
In this approach, the electronic Green's function $G$, whose poles yield the
single-particle excitation energies, is obtained by evaluating Feynman diagrams
representing many-electron interaction processes. Retaining only the lowest-order diagram
in an expansion in terms of the screened Coulomb interaction $W$ leads to the $GW$
method~\cite{Hedin65}. 
The $GW$ method greatly improves band gaps obtained from density-functional theory
(DFT)~\cite{Hybertsen85,Hybertsen86}, and further yields other accurate quasiparticle
properties, such as lifetimes and bandwidths~\cite{Rubio99,Jimenez97}, in a wide range of
weakly and moderately correlated materials.

However, despite its successes, the $GW$ method has well-known limitations.
Specifically,
it has proven difficult to systematically improve $GW$ theory by including higher-order
Feynman diagrams, so-called vertex corrections.
While extensions of the $GW$ approach have
been developed for specific applications -- such as the cumulant expansion of the
time-dependent Green's functions for the description of plasmon
satellites~\cite{Hedin80,Gunnarsson94,Aryasetiawan96} or the $T$-matrix approach for
magnetic systems~\cite{Springer98,Zhukov04,LischnerNa14} -- there exists currently no
universally accepted and applicable ``beyond-$GW$'' approach.
An additional problem in
most practical $GW$ calculations is a dependence of the results on the mean-field
starting point.
This arises because most implementations apply the $GW$ method as a
perturbative ``one-shot'' correction to a mean-field calculation, such as DFT or
Hartree-Fock (HF); this is usually referred to as the $G_0 W_0$ approach.
At a greater
numerical cost, self-consistent $GW$ calculations have been carried out with mixed
success~\cite{Shirley96,vonBarth96,Holm98,vanSchilfgaarde06,Caruso13}.

More common in \textit{ab initio} quantum chemistry, methods based on time-independent
many-body perturbation theory provide a different route to electronic excitations.
Importantly, in the time-independent framework, coupled-cluster theory provides a
well-studied and systematically improvable hierarchy within which to resum the
corresponding classes of Goldstone diagrams~\cite{Cizek66,ShavittBartlettBook,Bartlett07}.
Electronic excited states are obtained by equation-of-motion (EOM) coupled-cluster
theory~\cite{Monkhorst77,Stanton93,Krylov08}.
For molecules with weak to moderate correlations, coupled-cluster theories at the singles,
doubles, and perturbative triples level are established as the quantitative ``gold
standard'' of quantum chemistry~\cite{Bartlett07}.

While such \textit{ab initio} coupled-cluster theories have been widely applied to atoms
and molecules, they have traditionally been thought too expensive to use in extended
systems; for example, coupled-cluster theory with single and double excitations formally
has a computational scaling $O(N^6)$.
However, with improvements in algorithms and increases in computer power, the exciting
possibility of applying these methods to condensed matter problems is now within reach.
For example, very recent work has applied \textit{ground-state} coupled-cluster theory to
the uniform electron gas (UEG)~\cite{Shepherd13,Shepherd14} as well as atomistic
solids~\cite{Booth13}.
Correlated \textit{excited states} are the next frontier.

In this Letter, we apply, for the first time, EOM coupled-cluster theory to the UEG and
study its one-particle electronic excitations.
The UEG is a paradigmatic model of metallic condensed matter systems and these 
calculations illustrate
the potential of applying coupled-cluster theory in first-principles materials
simulations.
We employ coupled-cluster theory with single and double (and in some cases triple)
excitations; at this level, the diagrammatic content of our treatment goes significantly
beyond the standard $GW$ level of approximation.
As such, our coupled-cluster spectra allow us to assess the quality of vertex corrections
to the $GW$ method in the UEG.
For example, we evaluate the accuracy of the $GW$-plus-cumulant treatment of the
correlated satellite structure.
Further, as we consider the electron gas at the density of $r_\mathrm{s}=4.0$
corresponding to that of elemental sodium, our results for the occupied bandwidth provide
strong evidence to settle the long-standing puzzle concerning the interpretation of
photoemission experiments in this material.

\textit{Methods.}
We study electronic excitations of the three-dimensional UEG using a supercell approach,
i.e.~we place $N$ electrons in a cubic box of volume $\Omega$ with a neutralizing positive
background charge and periodic boundary conditions.
The thermodynamic limit is obtained, in principle, by increasing $N$ and $\Omega$ while
keeping the density $N/\Omega$ fixed.
Here, we only present results for the UEG with a Wigner-Seitz radius $r_\mathrm{s}=4.0$
($k_F=0.480$ a.u.) corresponding approximately to the valence electron density of metallic
sodium.
For the UEG Hamiltonian~\footnote{We treat the divergent $G=0$ component of the Coulomb
potential with the ``probe-charge'' Ewald summation method~\cite{Paier05}, 
i.e.~$v_{G=0} =\alpha_0/L$ where $\alpha_0 = 2.837\ 297\ 479$ is the Madelung constant 
of a 3D simple cubic lattice~\cite{Drummond08,Dabo08}.} 
we calculate the
one-electron Green's function
$G_\vk(\omega)$ and the corresponding spectral function $A_\vk(\omega) = \pi^{-1}
|\mathrm{Im} G_\vk(\omega)|$ using several methods: 
(i) mean-field theory, i.e.~HF and DFT in the local-density approximation
(LDA)~\cite{Perdew81}, 
(ii) time-dependent MBPT, i.e.~the $GW$ and $GW$+C methods,
(iii) EOM coupled-cluster theory, 
and (iv) dynamical density matrix renormalization group (DMRG), which provides
\textit{numerically exact} spectral functions for small system sizes~\cite{Jeckelmann02}; 
all DMRG calculations were performed with a bond dimension of $M=1000$.
Specifically, we compute spectral functions of occupied states, which are the ones probed
in photoemission experiments.

The one-particle eigenstates of the mean-field theories are plane-waves, $\phi_{\vk}(\vr)
= \Omega^{-1/2} e^{i\vk \cdot \vr}$.
These serve as a finite basis set, with a cutoff $k_\mathrm{cut}$, in the subsequent
MBPT, CC, and DMRG calculations.
The corresponding eigenenergies are given by $\epsilon_{\vk}=k^2/2 + V^{\mathrm{xc}}_{\vk}$, where
$V^{\mathrm{xc}}_{\vk}$ denotes the exchange-correlation matrix element, evaluated either at the HF
or DFT-LDA level (the Hartree term exactly cancels the interaction energy with the positive
background charge density). 

Based on the HF and DFT-LDA mean-field starting points, we carry out one-shot $GW$
(i.e.~$G_0W_0$) calculations~\cite{Hybertsen85,Hybertsen86} where screening is treated in
the random-phase approximation, as well as $G_0W_{\mathrm{xc}}$ calculations where screening is
treated with the DFT-LDA dielectric function~\cite{Northrup87,LischnerNa14}.
We also evaluate spectral functions using the $GW$-plus-cumulant (henceforth
$GW$+C) method.
This approximation yields the exact solution for a dispersionless core electron
interacting with plasmons~\cite{Langreth70} and noticeably improves the description of
plasmon satellite properties compared to $GW$, while retaining the accuracy of $GW$ for
the quasiparticle energies.
The $GW$+C formalism defines the Green's function as $G_{\vk}(t) =
G_{0,\vk}(t) \exp\left[-i\Sigma^\mathrm{x}_{\vk}t + C_{\vk}(t)\right]$, where $G_0$ is the 
Green's function from mean-field theory, $\Sigma^\mathrm{x}_{\vk}$ is the bare exchange 
self-energy and $C_{\vk}(t)=\pi^{-1} \int d\omega
|\text{Im}\Sigma_{\vk}(\omega+E^{GW}_{\vk})|(e^{-i\omega t}+i\omega t -1)/\omega^2$ is the
cumulant function~\cite{Hedin80,Gunnarsson94,Kas14}.
Here, $E^{GW}_{\vk}$ denotes the $GW$ orbital energy.
The $GW$+C approach has been applied to range of bulk
materials~\cite{Aryasetiawan96,Holm97,Guzzo11,Caruso15} and
nanosystems~\cite{Lischner13,Lischner14} and good agreement with experimental measurements 
on satellite structures was found.
However, comparisons of the $GW$+C to other accurate numerical calculations have been
difficult to perform, and this is one of the objectives below.

We perform EOM coupled-cluster calculations of the one-electron Green's function
starting from the mean-field ground-state determinant $|\Phi_0\rangle$, defined by the
occupied one-particle eigenstates with $k < k_F$. 
We briefly describe the relevant theory below; we refer to Ref.~\cite{ShavittBartlettBook}
for details. 
The coupled-cluster ground-state is defined as $|\Psi_0\rangle = e^T | \Phi_0\rangle$,
where the cluster operator is $T= \sum_{ia} t_i^a c^\dagger_a c_i +
\frac{1}{4}\sum_{ijab}t_{ij}^{ab} c^\dagger_a c^\dagger_b c_j c_i + ...$ (with the indices
$i,j$ referring to occupied states and the indices $a,b$ referring to unoccupied states). 
Singles, doubles, and triples coupled-cluster theories (denoted CCS, CCSD, and CCSDT)
correspond to truncating $T$ after one, two, and three electron-hole excitations. 
The $T$ operator and coupled-cluster ground-state energy are obtained through the
relations
\begin{equation}
\label{eq:cceqn}
\begin{split}
    E_0 &= \langle \Phi_0 |e^{-T} H e^T |\Phi_0\rangle 
        = \langle \Phi_0 | \bar{H} |\Phi_0\rangle \\
    0   &= \langle \Phi_i^a | \bar{H} |\Phi_0\rangle 
        = \langle \Phi_{ij}^{ab} | \bar{H} |\Phi_0\rangle = \dots,
\end{split}
\end{equation}
where the notation $\Phi_i^a$, $\Phi_{ij}^{ab}, \dots$ represents Slater determinants with
one, two, \dots\ electron-hole pairs, and $\bar{H}$ is the non-Hermitian coupled-cluster
effective Hamiltonian. 
By construction from Eq.~(\ref{eq:cceqn}), $|\Phi_0\rangle$ is the right ground-state
eigenvector of $\bar{H}$; its left ground-state eigenvector $\langle \tilde{\Phi}_0|$
takes the form $\langle \Phi_0| (1+ S)$, where $S = \sum_{ia} s_i^a c_a c^\dag_i +
\frac{1}{4}\sum_{ijab}s_{ij}^{ab} c_a c_b c^\dag_j c^\dag_i + ...$ creates excitations in
the bra, to the same level as in $T$.

Coupled-cluster excited states and energies are formally determined by diagonalizing
$\bar{H}$ in an appropriate space of excitations. 
For the single-particle (ionization) energies here, we diagonalize in the space of 1-hole
($1h)$ and 2-hole, 1-particle ($2h1p$) states for a CCSD ground-state, additionally
including the space of 3-hole, 2-particle ($3h2p$) states for a CCSDT
ground-state~\cite{Hirata00,Musial03}. 
The ionization contribution to the CC Green's
function~\cite{Nooijen92,Nooijen93} is then defined in the same space, as
\begin{align}
    G_{\vk}(\omega) = \langle \tilde{\Phi}_0| c^\dag_{\vk} P 
        \frac{1}{\omega - (E_0-\bar{H}) - i\eta} P c_{\vk} |\Phi_0\rangle
\end{align}
where $P$ projects onto the space of $1h$, $2h1p$, and (for CCSDT) $3h2p$ states.
We emphasize that although the initial ground-state CCSD calculation scales as $O(N^6)$,
the excited state ionization-potential EOM-CCSD has a reduced scaling $O(N^5)$; this 
should be compared to the $O(N^4)$ scaling of $GW$ methods.

\textit{Analysis of CC and GW methods}.
Coupled-cluster theory with $n$-fold electron-hole excitations in the $T$ operator
includes all time-independent diagrams with energy denominators that sum at most $n$
single-particle energies.
At the singles and doubles CCSD level (the lowest level used in this work), this already
includes more Feynman diagrams than are in $GW$ theory.
In particular, the CCSD energies and Green's function include not only the ring diagrams
which dominate the high-density limit of the electron gas~\cite{Gell-Mann57} and which
yield the screened RPA interaction in $GW$, but also ladder diagrams (such as generated in
$T$-matrix approximations) and self-energy insertions which couple the
two~\cite{Freeman77}.

Unlike $GW$ theory, CC approximations are invariant to the values of the single-particle
energies in the mean-field used to generate $|\Phi_0\rangle$.
They are further relatively insensitive to the single-particle orbitals, because $e^{T_1}$
parametrizes rotations from $|\Phi_0\rangle$ to any other determinant~\cite{Thouless60}.
While CC calculations typically start from a HF mean-field calculation, in the UEG the HF
and DFT mean-field theories share the same plane-wave states as their one-particle
eigenstates.
\textit{This means that the UEG CC calculations are completely invariant to the mean-field
choice (in the paramagnetic phase)}.
This complicates a fair comparison between one-shot $GW$ calculations and the CC
calculations.
For this reason, we present calculations with both HF (HF+$GW$ and LDA (LDA+$GW$) as a
reference; the former may be considered a fairer comparison with CC when assessing the
diagrammatic quality of the theories.

\textit{Results.} 
To establish the accuracy of the different methods, we initially study a supercell
containing 14 electrons in a minimal single-particle basis of 19 spatial orbitals
($k_\mathrm{cut}=0.572$ a.u.).
The electrons occupy seven orbitals, namely the orbital with $\vk=(0,0,0)$,
corresponding to the bottom of the band in the thermodynamic limit, and the six-fold
degenerate highest occupied orbital $\vk=(2\pi/L,0,0)$ corresponding to the Fermi
level in the thermodynamic limit.
For this small system, we can compare $GW$ and CCSD to coupled-cluster theory with all
triple excitations (CCSDT) as well as \textit{numerically exact} dynamical
density matrix renormalization group (DMRG) calculations of the spectral
function.

\begin{figure}[t] 
\includegraphics[scale=1.0]{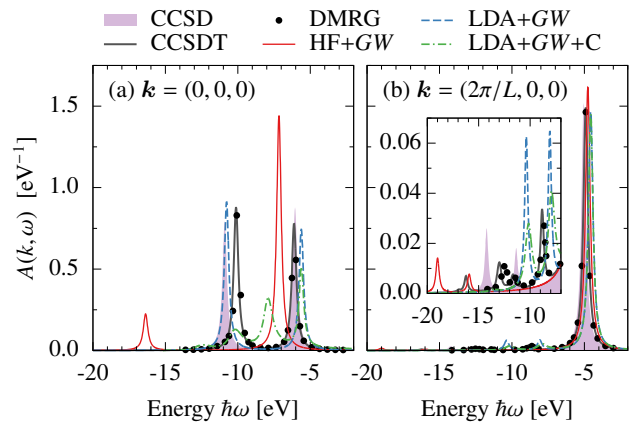} 
\caption{
    Spectral functions for the UEG with $r_\mathrm{s}=4.0$ using a supercell containing 14
    electrons in 19 spatial orbitals.
    (a) For the $\vk=(0,0,0)$ state, the spectral functions exhibits a prominent satellite
    peak.
    (b) For the highest occupied state at $\vk=(2\pi/L,0,0)$, the spectral function
    exhibits a strong quasiparticle peak with a very weak satellite structure; only
    converged data points are shown for DMRG. A linewidth broadening of $\eta=0.2$ eV is
    used in all calculations. 
} 
\label{fig:ueg1419} 
\end{figure}

Figure~\ref{fig:ueg1419}(a) shows our results for the deeply bound $\vk=(0,0,0)$ state.
All spectral functions (except for $GW$+C) exhibit two peaks: a quasiparticle peak near
$-6$~eV and a strong satellite peak near $-10$~eV nd excellent agreement between the CCSDT and the dynamical DMRG result.
The agreement between CCSD and the DMRG result is aso very good, in particular for the
quasiparticle peak.
Starting from the same HF reference as typically used in coupled-cluster theory, HF+$GW$ yields a
much less accurate result: the binding energy of the quasiparticle is too large by about 1
eV and the spectral weight is overestimated by almost a factor of 2.
This error is inherited from the underlying HF mean-field theory and illustrates the
starting point dependence of the method.
Even worse results are obtained for the satellite feature which is at far too low an
energy.
However, when starting from a DFT-LDA reference, the $GW$ approximation 
gives results with much improved accuracy, and is only slightly worse than CCSD.
(As discussed above, in the UEG the CC results are invariant to the reference).

Interestingly, $GW$+C yields several satellite peaks with incorrect energies and
underestimated peak heights, illustrating some of the challenges in systematically
improving on $GW$ theory through standard vertex corrections.
By construction, the $GW$+C approach produces a plasmon-replica satellite structure (see
below) even for small systems, which is physically incorrect.

Consistent with Fermi liquid theory, the spectral functions of the $\vk=(2\pi/L,0,0)$
state shown in Fig.~\ref{fig:ueg1419}(b) exhibit significantly weaker electron
correlations than the spectral functions of the $\vk=(0,0,0)$ state.
Specifically, all methods predict a strong quasiparticle peak with a binding energy of
about 5 eV and weak satellite features.
The inset of Fig.~\ref{fig:ueg1419}(b) shows that the detailed structure of the satellites
is quite complex.
While CCSDT accurately captures the complex features seen in the exact spectrum, none of the other
methods are fully satisfactory.
In particular, HF+$GW$ pushes satellite features to too low energies, the LDA+$GW$ places
the satellite peaks at too high an energy, and CCSD places them in between.
$GW$+C correctly reduces the weight of the main $GW$ satellite peaks but does not
otherwise improve the spectrum.

Next, to study the approach to the thermodynamic limit, we carried out calculations
on larger supercells for which CCSDT and dynamical DMRG are no longer computationally
tractable.
We performed CCSD, $GW$, and $GW$+C calculations for supercells containing 38, 54, 66
and 114 electrons; here we will only discuss the largest system studied.
For the 114 electron system, we used plane-wave basis sets with at least 485 spatial
orbitals, corresponding to $k_\mathrm{cut}=0.985$ a.u, which is sufficiently large to
converge all peak positions to within 0.2~eV.

\begin{figure}[t] 
\includegraphics[scale=1.0]{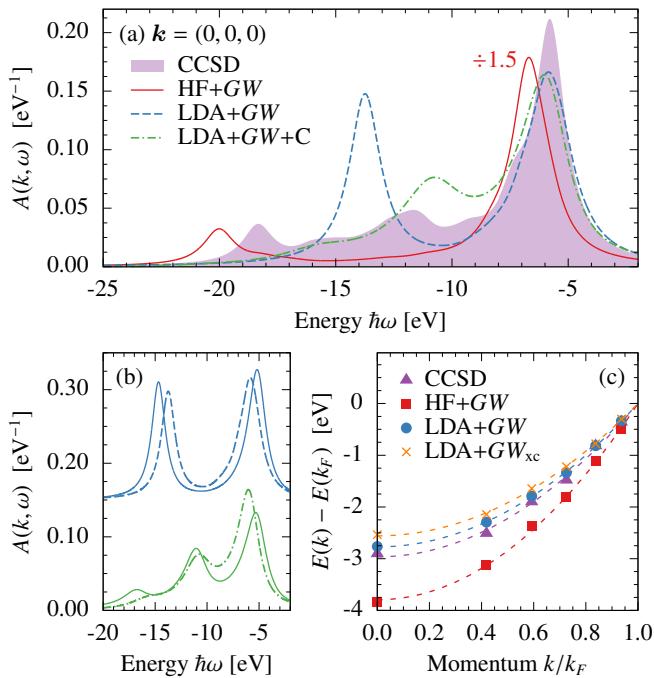} 
\caption{
    (a) Spectral function of the $\vk=(0,0,0)$ state of the 3D UEG with $r_\mathrm{s}=4.0$
    and 114 electrons in 485 orbitals.
    The HF+$GW$ result is scaled down by a factor of 1.5 and a linewidth broadening of
    $\eta=0.8$ eV is used in all calculations.
    (b) Comparison of the spectral functions of the $\vk=(0,0,0)$ state in the
    thermodynamic limit (solid curves) and the 114 electron system (dashed curves) from
    LDA+$GW$ (blue curves) and LDA+$GW$+C (green curves).
    (c) Complete basis set limit quasiparticle energies as a function of wave vector for
    the 114 electron system (symbols) and quadratic fits (dashed curves).
} 
\label{fig:ueg114} 
\end{figure}

Figure~\ref{fig:ueg114}(a) shows the spectral function of the $\vk=(0,0,0)$ state for the
UEG with 114 electrons in 485 orbitals.
The CCSD spectral function exhibits a strong quasiparticle peak near $-6$ eV.
For the $GW$ calculations, we observe again a strong dependence on the mean-field starting
point: while the quasiparticle energy from LDA+$GW$ agrees very well with CCSD, that from
HF+$GW$ is significantly worse.
This is not surprising since DFT-LDA yields much more accurate metallic bands than HF.

At higher binding energies, the CCSD spectral function exhibits a rather complex satellite
structure, however two major regions of spectral weight can be identified near $-12$ eV
and $-18$ eV.
In contrast, both the HF+$GW$ and the LDA+$GW$ spectral functions exhibit only a single,
prominent satellite peak.
Lundqvist and co-authors~\cite{Hedin67,Lundqvist67} assigned this peak to a novel excited
state, the plasmaron, which is a coherent superposition of strongly coupled plasmon-hole
pairs.
While several experiments reported the observation of plasmaron excitations in doped
graphene and semiconductor quantum-well two-dimensional electron gases, it has recently
become clear that their prediction by $GW$ is \textit{spurious}.
Vertex-corrected time-dependent MBPT approaches, such as the $GW$+C method, do not predict
such a state and instead yield a satellite structure that consists of an infinite series
of peaks corresponding to the ``shake-up'' of one or more
plasmons~\cite{Langreth70,Hedin80}.
Notably, the major peaks in the CCSD spectral function are separated by roughly 6~eV
corresponding to the classical plasma frequency $\omega_\mathrm{P}=5.9$~eV in an electron
gas with $r_\mathrm{s}=4.0$.
Comparing the LDA+$GW$+C result to CCSD in Fig.~\ref{fig:ueg114}(a), we find a
qualitatively similar spectrum.
However, at least at this system size, the CCSD spectral function has a stronger
quasiparticle peak, a larger spectral width, and significantly more fine-structure than
the $GW$+C spectral function.

To assess remaining errors of the 114 electron system relative to the thermodynamic limit,
we compare the $\vk=(0,0,0)$ spectral functions of the UEG with 114
electrons with the results fully converged to the thermodynamic limit for the LDA+$GW$ and
the LDA+$GW$+C methods. Fig.~\ref{fig:ueg114}(b) shows good qualitative agreement between 
the two sets of spectral functions for this class of methods.

Finally, Fig.~\ref{fig:ueg114}(c) shows the quasiparticle energies as function of the
electron wave vector, i.e.~the energy dispersion relation, for the 114 electron
system~\footnote{For this data, we performed an extrapolation to the complete basis set
(CBS) limit. 
For each calculation with $M$ basis functions, the quasiparticle peaks
were fitted with a Lorentzian lineshape.
These peak positions were observed to have a $1/M$ dependence, allowing for extrapolation
to $M\rightarrow\infty$.
As a function of wavevector, we then fitted the CBS-limit peak positions to a quadratic
``effective mass'' dispersion and referenced all energies to the extrapolated Fermi energy
in the thermodynamic limit.}.
The inferred bandwidths are 2.96~eV for CCSD, 3.79~eV for HF+$GW$, 2.77~eV for LDA+$GW$,
and 2.56~eV for LDA+$GW_{\mathrm{xc}}$.
While DFT-LDA gives a bandwidth of 3.13~eV, HF predicts a value of 7.29~eV, significantly
larger than any other method.
The failure of HF to describe metallic systems is well-documented and results from the
absence of screening.

The bandwidth of simple metals, and in particular sodium, has been the subject of a
decades-long debate.
Plummer and co-workers~\cite{Jensen85,Lyo88} carried out angle-resolved photoemission
experiments on sodium and reported a bandwidth of 2.5--2.65~eV, significantly smaller than
the free-electron and DFT-LDA value of $\sim$3.1~eV, and even the LDA+$GW$ value of
$\sim$2.8~eV~\cite{Hedin65}.
Interestingly, the experimental result agrees quite well with the bandwidth from a
LDA+$GW_{\mathrm{xc}}$ calculation~\cite{Northrup87,LischnerNa14}, which contains vertex
corrections for the dielectric function; however, including vertex corrections also in
the self-energy increases the bandwidth again~\cite{Yasuhara99,Mahan00,Stankovski07}.
As an alternative explanation, Shung and Mahan~\cite{Shung86,Shung87} suggested that the
measured bandwidth results from many-body effects in combination with final-state effects
and an interference between surface and bulk photoemission.
The close agreement seen here between the quasiparticle dispersion of LDA+$GW$ and CCSD --
especially the \textit{larger} bandwidth of CCSD -- suggests that the theoretical
description of the quasiparticle peak positions may be adequate already 
and supports Shung and Mahan's thesis that the remaining discrepancy in the observed 
bandwidth is due to final-state and interference effects.

\textit{Conclusion.} 
We have demonstrated the first application of coupled-cluster techniques to the
computation of spectra in condensed phase systems, using the uniform electron gas as a
model system.
For finite uniform electron gas models of various sizes we find that coupled-cluster, even
at the singles and doubles level (CCSD), provides improvement over $GW$ and
even $GW$-plus-cumulant theory. 
Interestingly, while the latter exhibits good accuracy for large systems (producing
reasonable plasmon-like satellite structures), the
former is significantly more accurate for small systems;
CCSD naturally interpolates between these two limits.
In conclusion, by providing a systematic framework that goes beyond the diagrammatic
content of the $GW$ approximation, coupled-cluster theories represent 
a very promising, new direction
in the search for more accurate methods to compute the spectra of real materials.

\vspace{1em}
\acknowledgments{
CCSD calculations were carried out using a modified version of the \textsc{ACES III}
code~\cite{Lotrich08} through the University of Florida High Performance Computing Center.
CCSDT calculations were performed using a modified version of the \textsc{CFOUR}
code~\cite{CFOUR}.
Dynamical DMRG calculations were done with the \textsc{BLOCK}
code~\cite{Sharma12,Olivares-Amaya15,Dorando09}.
J.L.~acknowledges support from EPSRC under Grant No.~EP/N005244/1 and also from the Thomas
Young Centre under Grant No.~TYC-101.  S.G.L.~and J.L.~acknowledge supported by the SciDAC
Program on Excited State Phenomena in Energy Materials funded by the U. S. Department of
Energy (DOE), Office of Basic Energy Sciences and of Advanced Scientific Computing
Research, under Contract No.~DE-AC02-05CH11231 at Lawrence Berkeley National Laboratory
(algorithm and code development) and by the National Science Foundation under grant
DMR15-1508412 (basic theory and formalism).
D.A.M.~is supported at the University of Texas by the National Science Foundation
(ACI-1148125).
E.R.~acknowledges the Computational Laboratory for Hybrid/Organic Photovoltaics of
CNR-ISTM for a fellowship funded by the CNR-EFOR project.
T.C.B.~is supported by the Princeton Center for Theoretical Science. 
J.M., T.W., E.R., and G.K.-L.C.~acknowledge primary support from DOE (SciDAC): Predictive
Computing for Condensed Matter under contract no.~DE-SC0008624, and additional funding
from DOE: Quantum Embedding for Correlated Electronic Structure in Large Systems and the
Condensed Phase under contract no.~DE-SC0010530.
G.K.-L.C.~also acknowledges support from the Simons Foundation through the Simons
Collaboration on the Many Electron Problem.
}

\end{document}